\documentclass[
 aip,
 amsmath,amssymb,
 reprint,%
floatfix,
]{revtex4-1}

\usepackage{graphicx}
\usepackage{placeins} 
\usepackage{float}
\usepackage{wrapfig}
\usepackage{dcolumn}
\usepackage{bm}
\usepackage[dvipsnames]{xcolor}
\usepackage[mathlines]{lineno}

\usepackage[utf8]{inputenc}
\usepackage[T1]{fontenc}
\usepackage{mathptmx}
\usepackage[english]{babel}
\usepackage{wasysym}
\usepackage{fdsymbol}

\makeatletter
\def\bbl@set@language#1{%
  \edef\languagename{%
    \ifnum\escapechar=\expandafter`\string#1\@empty
    \else\string#1\@empty\fi}%
  \@ifundefined{babel@language@alias@\languagename}{}{%
    \edef\languagename{\@nameuse{babel@language@alias@\languagename}}%
  }%
  \select@language{\languagename}%
  \expandafter\ifx\csname date\languagename\endcsname\relax\else
    \if@filesw
      \protected@write\@auxout{}{\string\select@language{\languagename}}%
      \bbl@for\bbl@tempa\BabelContentsFiles{%
        \addtocontents{\bbl@tempa}{\xstring\select@language{\languagename}}}%
      \bbl@usehooks{write}{}%
    \fi
  \fi}
\newcommand{\DeclareLanguageAlias}[2]{%
  \global\@namedef{babel@language@alias@#1}{#2}%
}
\makeatother

\DeclareLanguageAlias{en}{english}

\begin{document}

\preprint{AIP/123-QED}


\title{Solid-State Reactions at Niobium-Germanium Interfaces in Hybrid Superconductor-Semiconductor Devices}

\author{B. Langa Jr.}
\affiliation{%
Department of Physics and Astronomy, Clemson University, Clemson, South Carolina 29634, USA
}%

\author{D. Sapkota}
\affiliation{%
Department of Physics and Astronomy, Clemson University, Clemson, South Carolina 29634, USA
}%
\affiliation{ 
NASA Goddard Space Flight Center, Greenbelt, Maryland 20771, USA
}%

\author{I. Lainez}
\affiliation{%
Department of Physics and Astronomy, Clemson University, Clemson, South Carolina 29634, USA
}%

\author{R. Haight}
\affiliation{%
IBM TJ Watson Research Center, Yorktown Heights, New York 10598, USA
}%

\author{B. Srijanto}
\affiliation{%
Center for Nanophase Materials, Oak Ridge National Laboratory, Oak Ridge, Tennessee 37830, USA
}%

\author{L. Feldman}
\affiliation{%
Department of Physics and Astronomy, Rutgers University, Newark, New Jersey 08901, USA
}%

\author{H. Hijazi}
\affiliation{%
Department of Physics and Astronomy, Rutgers University, Newark, New Jersey 08901, USA
}%

\author{X. Zhu}
\affiliation{%
Department of Materials Science and Engineering, University of Texas in Dallas, Richardson, Texas 75080, USA
}%

\author{L. Hu}
\affiliation{%
Department of Materials Science and Engineering, University of Texas in Dallas, Richardson, Texas 75080, USA
}%

\author{M. Kim}
\affiliation{%
Department of Materials Science and Engineering, University of Texas in Dallas, Richardson, Texas 75080, USA
}%

\author{K. Sardashti}
\affiliation{%
Department of Physics and Astronomy, Clemson University, Clemson, South Carolina 29634, USA
}%
\affiliation{ 
Department of Physics, University of Maryland, College Park, Maryland 20740, USA
}%

\date{\today}

\begin{abstract}
Hybrid Superconductor-Semiconductor (S-Sm) materials systems are promising candidates for quantum computing applications. Their integration into superconducting electronics has enabled on-demand voltage tunability at millikelvin temperatures. Ge quantum wells (Ge QWs) have been among the semiconducting platforms interfaced with superconducting Al to realize voltage tunable Josephson junctions. Here, we explore Nb as a superconducting material in direct contact with Ge channels by focusing on the solid-state reactions at the Nb/Ge interfaces. We employ Nb evaporation at cryogenic temperatures ($\sim$100 K) to establish a baseline structure with atomically and chemically abrupt Nb/Ge interfaces.  By conducting systematic photoelectron spectroscopy and transport measurements on Nb/Ge samples across varying annealing temperatures, we elucidated the influence of Ge out-diffusion on the ultimate performance of superconducting electronics. This study underlines the need for low-temperature growth to minimize chemical intermixing and band bending at the Nb/Ge interfaces.
\end{abstract}

\maketitle


In the past decade, research on hybrid superconducting-semiconducting (S-Sm) materials and devices has focused on realizing functionalities that enhance the operation of superconducting quantum devices. Hybrid S-Sm-S Josephson junctions (JJs) have been at the core of such efforts being integrated into devices such as voltage-tunable qubits (Gatemons) and voltage-tunable couplers.~\cite{sardashti_voltage-tunable_2020, casparis_voltage-controlled_2019, strickland_superconducting_2023, larsen_semiconductor-nanowire-based_2015,casparis_superconducting_2018} S-Sm-S JJs rely on superconductivity induced within a semiconducting channel through Andreev reflection.~\cite{klapwijk_proximity_2004} This configuration simultaneously benefits from superconducting properties (i.e., dissipationless charge transport) and semiconducting properties (i.e., tunable carrier densities, long mean free paths, large spin-orbit coupling). 

Hybrid S-Sm systems composed of superconducting Al contacts coupled to InAs quantum wells (QWs) have been extensively studied as a platform for hybrid S-Sm quantum devices.~\cite{shabani_two-dimensional_2016, wickramasinghe_transport_2018} However, there are concerns with the high losses in the InAs heterostructures at microwave frequencies, as evidenced by low internal quality factors in superconducting resonators and qubits fabricated on the Al-InAs platforms.~\cite{strickland_superconducting_2023, casparis_superconducting_2018, kringhoj_magnetic-field-compatible_2021, oconnell_yuan_epitaxial_2021, strickland_characterizing_2024} Current performance limitations have been attributed to the large defect densities within the heterostructures and the piezoelectricity in indium phosphide (InP) substrates.~\cite{strickland_superconducting_2023} Therefore, exploring alternative materials systems is critical to realizing coherent gate-tunable quantum devices.

\begin{figure*}[ht!]
\includegraphics[width=\textwidth]{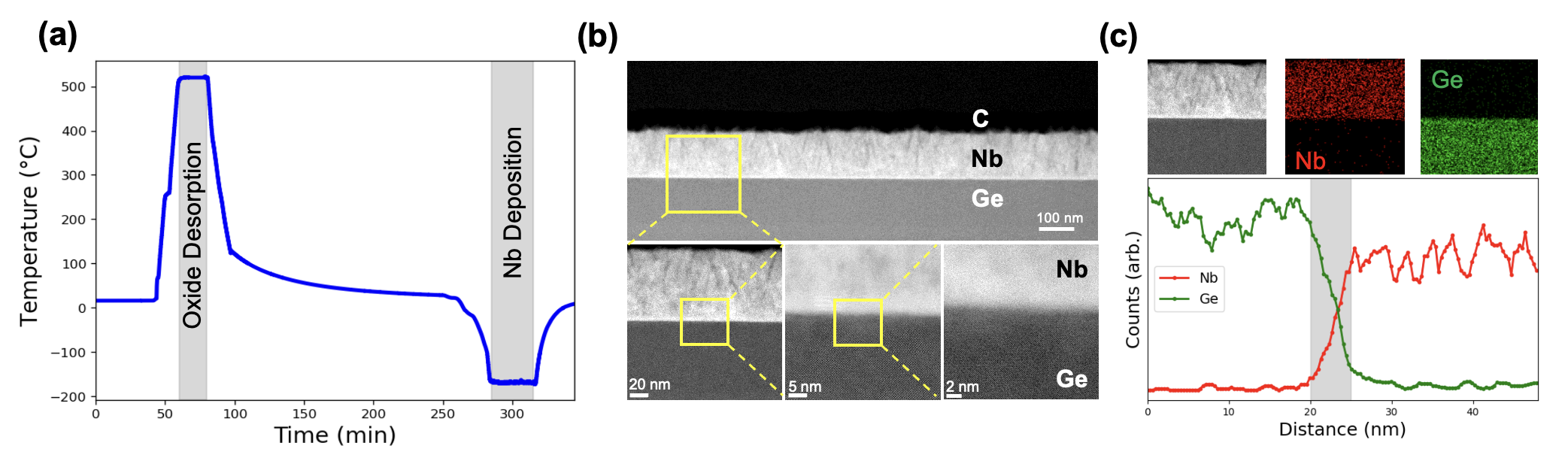}
\caption{\label{fig1} (a) Substrate temperature as a function of time during a typical cryogenic Nb growth process starting with GeO$_2$ desorption at 500 $^{\circ}$C and cooling the sample to -170 $^{\circ}$C for the Nb e-beam deposition. (b) Cross-sectional TEM image of a cryogenically grown 100 nm thick film of Nb on Ge(001) showing a nanocrystalline structure with grain sizes ranging from 5 to 10 nm. (c) Elemental analysis of the Nb/Ge(001) interface by EDS showing minimal intermixing and a narrow interface layer  ( < 4 nm).}
\end{figure*}

Ge QWs have recently emerged as promising alternatives to InAs QWs due to their high hole mobility, the possibility for growth on Si substrates with low microwave loss and compatibility with wafer-scale fabrication processes. Several studies have demonstrated voltage-tunable hybrid Al-Ge QW JJs with critical currents ranging from 6 to 500 nA.~\cite{vigneau_germanium_2019, tosato_hard_2023} Al has been the sole superconducting contact layer in those studies as it forms a stable ohmic contact to the Ge QWs~\cite{vigneau_germanium_2019, aggarwal_enhancement_2021} Integrating superconducting materials with larger gaps, such as Nb, may be of interest for realizing a more robust induced superconducting gap in the Ge QW devices. Thus far, studies of hybrid Nb/Ge QW systems have been hindered by two challenges: 1) sputtering, which is conventionally used for Nb deposition and imparts significant physical damage to the substrate during the deposition; 2) solid-state chemical reactions between Nb and Ge under vacuum and at elevated temperatures are only studied over limited temperature windows primarily above 900 $^{\circ}$C.~\cite{iijima_reaction_1979} 

In this study, we leverage a cryogenic ultrahigh vacuum (UHV) e-beam evaporation process to directly deposit thin films of Nb on Ge substrates. Using this method we demonstrate chemically abrupt and atomically sharp Nb/Ge interfaces with near-Ohmic characteristics (barrier height $\sim$121 meV). We study thermally-driven changes in the physical and chemical properties of the Nb/Ge heterostructures by annealing the UHV-grown films at a wide range of temperatures from 150 $^{\circ}$C to 675 $^{\circ}$C. With increased temperature, we observed a continuous rise in the Ge content of the Nb films reaching 65\% at 675 $^{\circ}$C.  This is accompanied by an increase in the barrier height at the Nb/Ge interface up to 575 $^{\circ}$C followed by a sharp drop due to the significant incorporation of Ge. The superconducting properties of the films show a slight improvement after a 300 $^{\circ}$C anneal mainly due to structural refinements in the Nb films. Further annealing led to degradation of the superconducting properties.

To prepare the Nb/Ge heterostructures we started with undoped Ge(001) wafers ($\rho$ > 50 $\Omega$.cm). The Ge(001) substrates were cleaned by a 5-min sonication in acetone, isopropanol, and deionized water followed by N$_2$ drying. The Ge substrates were then loaded into a UHV molecular beam epitaxy (MBE) system and annealed at 150 $^{\circ}$C for 30 minutes before introduction into the growth chamber. With pressures less than 1 x 10$^{-9}$ mbar, the MBE allows for the deposition of high-quality Nb thin film. The MBE system is equipped with a manipulator capable of cooling the sample to cryogenic temperatures (T < 200 K) to minimize intermixing between the Nb and Ge. Throughout the deposition, we used a reflective high-energy electron diffraction (RHEED) to monitor the surface reconstruction of the Ge and Nb film. Figure 1(a) depicts a typical substrate temperature profile for the Nb growth process. It begins with Ge oxide desorption at 500 $^{\circ}$C, followed by cooling to approximately -170 $^{\circ}$C for the deposition of the Nb thin films. The Nb thin films, varying in thickness from 8 nm to 200 nm, were deposited using a vertical electron gun at rates of 0.8 –1.1 Å/s.

We examined the nanoscale structure of the cryogenically-grown Nb/Ge heterostructure with cross-sectional high-angle annular dark field (HAADF) TEM imaging (JEOL-ARM 200F operated at 200 kV). The cross-sectional STEM specimen was prepared using an FEI Nova 200 dual-beam focused ion beam. Based on the HAADF images, the as-grown 100 nm thick Nb film is nanocrystalline with an average grain size of 8 nm (see Fig. 1 (b)).  Figure 1(c) shows energy-dispersive X-ray spectroscopy (EDS) maps and line scans across a narrow Nb/Ge interface region. The EDS results confirm that the growth at -170 $^{\circ}$C leads to a chemically abrupt interface with minimal intermixing and a narrow interface layer (< 4 nm) between the Nb and Ge. For our TEM sample thickness of 50 nm, the beam broadening is estimated to be 4.07 nm confirming an atomically sharp Nb/Ge interface.   This abruptness provides a baseline for testing the influence of thermal processing on the physical and chemical properties of the Nb/Ge heterostructures.

To explore the changes in the surface and interface electronics we use femtosecond ultraviolet photoelectron spectroscopy (fs-UPS) (See supplementary material for fs-UPS details). ~\cite{lim_oxygen_2005, lim_situ_2005, lim_temperature_2005} Figure 2(a) shows the valence band spectra measured by UPS for a 30 nm thick Nb film deposited on Ge(001) grown at -170 $^{\circ}$C as a function of in-situ UHV annealing at temperatures ranging from 150 $^{\circ}$C to 675 $^{\circ}$C. After the samples were grown, they were stored in a stainless steel canister and pressurized to 2.0 bar with ultrahigh-purity N$_2$ to avoid any oxidation during the transfer to the fs-UPS lab. The samples were then immediately loaded into the UHV system. As the sample temperature increases, the valence band edge becomes progressively sharper, consistent with a larger density of states at the Fermi level and a good metallic behavior. Additionally, above 625 $^{\circ}$C, new peaks emerge at higher binding energies (BE = 6.5 and 11.7 eV) that may be due to Ge incorporation into the Nb film.

\begin{figure*}[ht]
\includegraphics[width=0.7\textwidth]{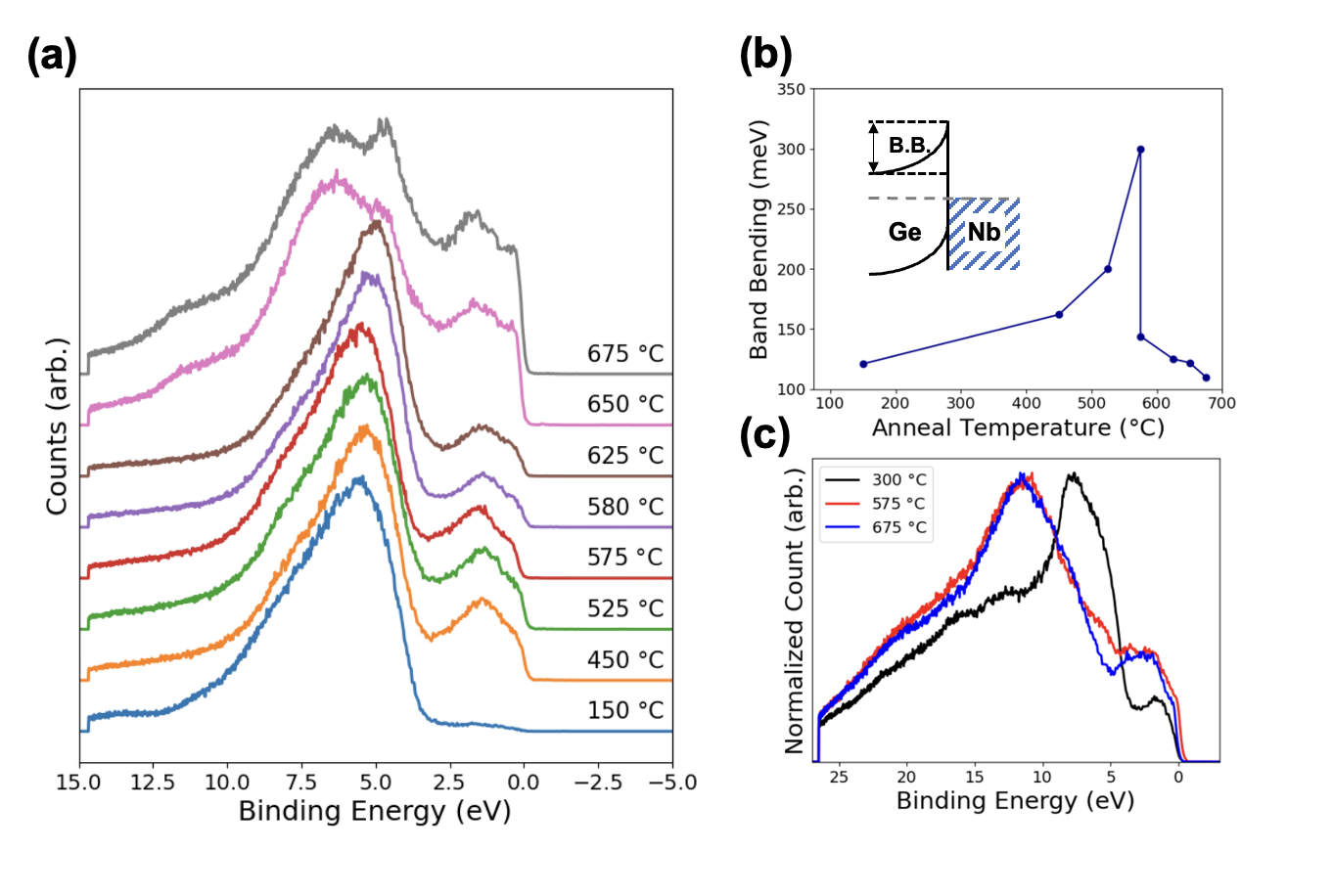}
\caption{\label{fig2} (a) Valence band spectra vs binding energy for different annealing temperatures of a 30 nm Nb film on Ge(001). Higher annealing temperatures show sharper valence bands at the Fermi level edge. (b) Band bending at the Nb/Ge interface as a function of annealing temperature for a 30 nm Nb film. (c) Valence band spectra for 200 nm thick Nb films annealed at 300 $^{\circ}$C, 575 $^{\circ}$C, and 675 $^{\circ}$C.}
\end{figure*}

Figure 2(b) shows the band bending at the Nb/Ge interface as a function of anneal temperature through a pump and probe process. To determine the band bending at the Nb/Ge interface, synchronized 1.55 eV pump pulses are concentrated on the probe area. This action generates an electron-hole pair within the interface, effectively counteracting the static dipole field within the depletion region, thus flattening the energy bands. Consequently, the UPS spectrum undergoes a consistent shift in energy. The extent of band bending is determined by measuring this shift. Initially, the sample exhibits a relatively low band bending of 121 meV at 150 $^{\circ}$C, which progressively rises with temperature to 300 meV at 575 $^{\circ}$C. This is followed by a sharp fall in the band bending to 144 meV upon longer anneal at 575 $^{\circ}$C. With further increase in temperature, band bending continuously decreases to 110 meV at 675 $^{\circ}$C. This decrease in the Schottky barrier height may be attributed to the degrading chemical abruptness of Nb/Ge interfaces at elevated temperatures consistent with the increased work function of our annealed Nb films (see Supplementary Material, Fig. S1).

To confirm that the additional peaks in the valence band spectra are due to Ge out-diffusion in Nb, we repeated our fs-UPS measurements on 200 nm thick Nb films on Ge (001) as shown in Fig. 2(c). Similar to thinner Nb films, high binding energy peaks emerged in the valence band spectra when 200 nm thick Nb films were annealed at 575 $^{\circ}$C and 675 $^{\circ}$C. Nevertheless, the peak at 11.7 eV exhibited a greater intensity in thicker films compared to the peak at 6.5 eV. Moreover, the annealing led to a sharper valence band edge on the 200 nm thick films. The sharp valence bands and lower band bending are indicators of an Nb/Ge interface that may be suitable for hybrid S-Sm JJs. However, the influence of Ge incorporation during the annealing process on surface and interface electronics could potentially restrict its beneficial effects.

To investigate the impact of annealing on the composition of the Nb films, we performed compositional analysis on the Nb/Ge heterostructures using ex-situ X-ray photoelectron spectroscopy (XPS). The data was collected using a PHI VersaProbe III with a monochromatic Al Ka X-ray source (hv=1486.6 eV). (See supplementary material for details). Figures 3(a) and 3(b) show the Nb3p and Ge2p spectra for the Nb/Ge samples as a function of annealing temperature. The samples were annealed ex-situ in the UHV MBE system and were transferred to the XPS lab under N$_2$. They were immediately loaded into the XPS system and underwent 3 cycles of Ar sputtering at 2 kV over a 3 x 3 mm$^2$ area. For the as-grown sample (black traces), no Ge signal is detected while the Nb3p peaks appear at the expected binding energies for elemental Nb (BE$_{\text{Nb3/2}}$ = 360.7).~\cite{sasaki_chemical-state_1985} On the other hand, annealing the sample to 300 $^{\circ}$C shows the emergence of Ge on the top surface. The Ge2p peaks continue shifting to higher binding energies with temperature eventually aligning with the expected binding energies of elemental Ge2p peaks (BE$_{\text{Ge3/2}}$ = 1217.4).~\cite{morgan_binding_1973} Similarly the Nb3p peaks shift to higher BEs by increasing the annealing temperature. This is consistent with a notable accumulation of Ge near the top surfaces of Nb films, with a fraction of the Ge atoms bonding to Nb atoms. 

\begin{figure*}[t]
\includegraphics[width=\textwidth]{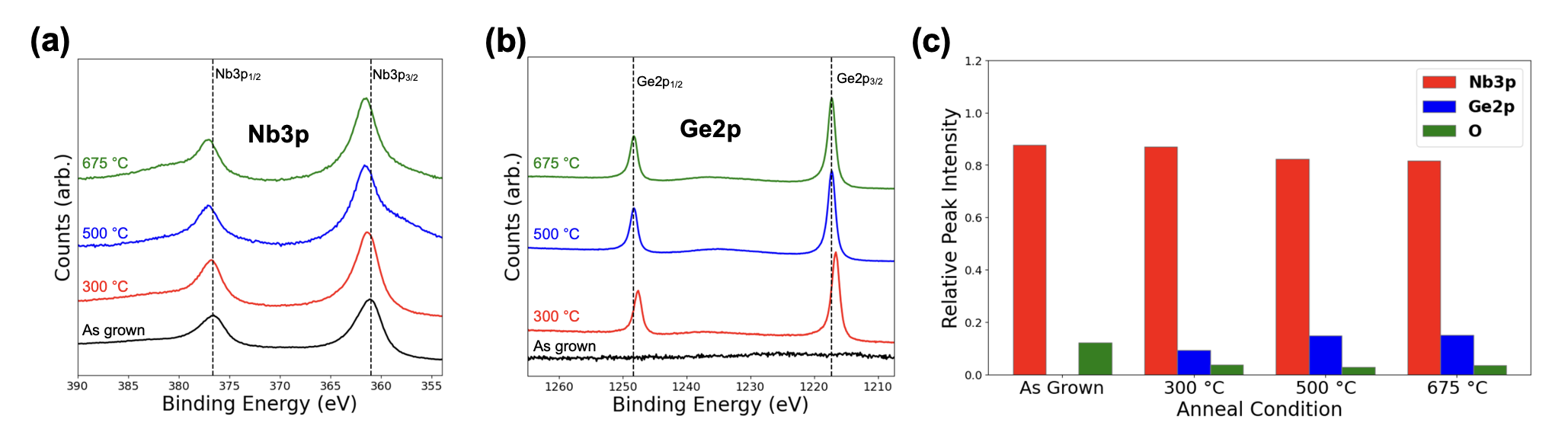}
\caption{\label{fig3} (a) Anneal temperature dependence of Nb3p peaks. Peaks shift to higher binding energies as anneal temperature increases. (b) Anneal temperature dependence of Ge2p peaks. Ge2p peaks increase in signal the higher the anneal temperature. (c) Overall elemental analysis of each sample.}
\end{figure*}

Figure 3(c) displays the elemental composition of the Nb films derived from the Nb3p, Ge2p, and O1 peaks.  To calculate the intensity of each peak, its area was first corrected by its relative sensitivity factor (RSF). The intensity of each element was then divided by the total sum of intensities across all three peaks to yield the elemental ratios. The as-grown sample started with Nb:Ge of 1:0. After annealing at 300 $^{\circ}$C, Nb:Ge changed to 1:0.11. The final two anneal temperatures (500 $^{\circ}$C and 675 $^{\circ}$C) shows further Ge incorporation with Nb:Ge of 1:0.18. Considering the small inelastic mean free paths~\cite{tanuma_calculations_1994} for the Nb3p (1.07 nm) and Ge2p (0.40 nm) photoelectrons, our results point to sizable Ge accumulation on the top surfaces of Nb even at temperatures as low as 300 $^{\circ}$C.

\begin{figure*}[ht]
\centering
\includegraphics[width=\textwidth]{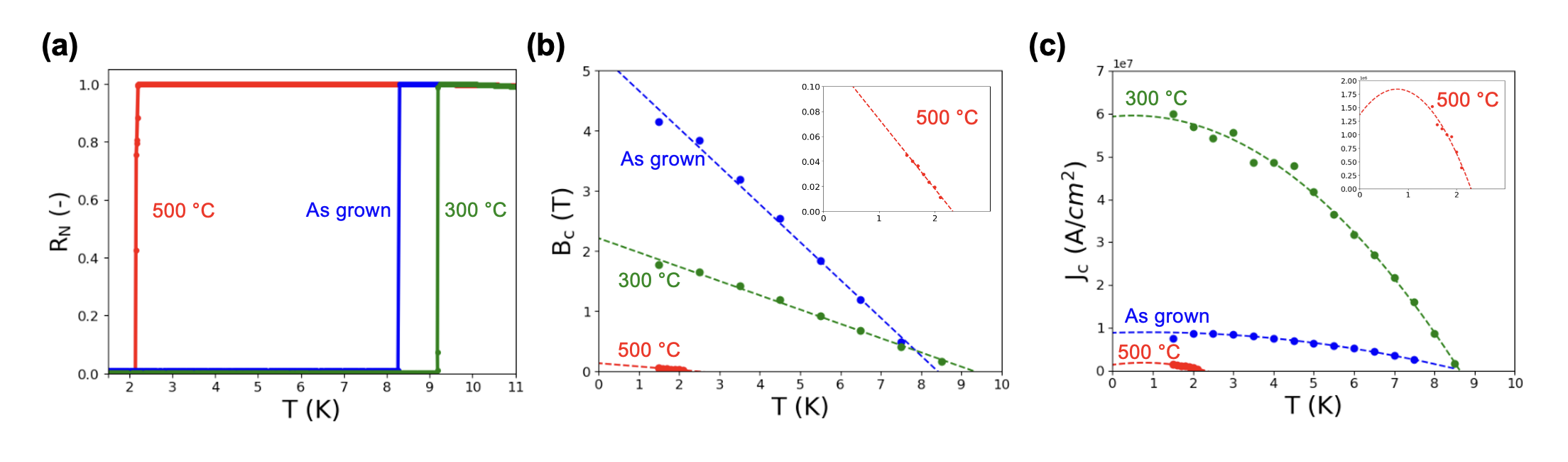} 
\caption{\label{fig4} Superconducting transport properties of Nb/Ge heterostructure as grown (blue) and after UHV annealing at 300 $^{\circ}$C (green) and 500 $^{\circ}$C (red). (a) Normalized resistance (R$_\text{N}$) as a function of temperature. (b) Dependence of critical magnetic field (B$_\text{C}$) on temperature. (c) Critical current density (J$_\text{C}$) as a function of temperature.}
\end{figure*}

To complement the surface composition, derived from XPS, with bulk film composition, we utilized Rutherford Backscattering Spectrometry (RBS). Three 30 nm thick Nb films were studied by RBS: 1) as-grown; 2) annealed at 585 $^{\circ}$C 3) annealed at 675 $^{\circ}$C. The samples were annealed in situ in UHV and transferred to the RBS system in a stainless steel canister filled with 2.0 bar of ultrahigh-purity N$_2$ to prevent oxidation. Table 1 summarizes the RBS compositional analysis results for the three Nb/Ge samples. The as-grown sample show no Ge incorporation consistent with XPS results. On the other hand, the two samples annealed at 585 $^{\circ}$C and 675 $^{\circ}$C show a 1:1 and 1:1.85 of Nb:Ge, respectively. These results are in agreement with the cross sectional Stem-EDS measuremtns on the films where significant Ge diffusion was observed at 650 $^{\circ}$C. (see Supplementary Material, Fig. S2). Moreover Ge diffusion resulted in  increased grain size and surface roughness in the Nb films. (See Supplementary Material, Fig. S3). We anticipate the presence of Ge on the top surface and within the bulk to change the superconducting properties of the Nb films.

\begin{table}[hb]
\caption{\label{tab:table1}Summary of the RBS results on three Nb/Ge samples with different thermal processing conditions from as-grown to annealed at 585 $^{\circ}$C and 675 $^{\circ}$C.}
\begin{ruledtabular}
\begin{tabular}{cccc}
Anneal Temperature&Nb (10$^{15}$ cm$^{-2})$&Ge (10$^{15}$ cm$^{-2})$&Nb:Ge\\
\hline
As-grown & 130$\pm$5 & - & 1:0 \\
585 $^{\circ}$C & 125$\pm$5 & 125$\pm$5 & 1:1 \\
675 $^{\circ}$C & 125$\pm$5 & 230$\pm$5 & 1:1.85 \\
\end{tabular}
\end{ruledtabular}
\end{table}

To evaluate the influence of thermal processing on the superconducting properties of the Nb/Ge heterostructures we utilized a cryogenic measurement system (Oxford Instruments Teslatron PT).  For a detailed evaluation of superconducting parameters such as superconducting transition temperature (T$_\text{C}$), critical magnetic field (B$_\text{C}$), and critical current density (J$_\text{C}$), Nb thin films were fabricated into microwires of 100 $\mu$m length and varying widths (5, 10, and 20 $\mu$m).  Details of the microfabrication process are provided in the supplementary material. Figure 4 displays the transport measurement results for 10 $\mu$m wide microwires fabricated on 100 nm thick Nb films that were as-grown (blue), annealed at 300 $^{\circ}$C (green), and annealed at 500 $^{\circ}$C (red). The annealing was conducted on the thin films in UHV (5 x 10$^{-10}$ mbar) immediately after the cryogenic growth. Table 1 lists the key superconducting parameters extracted from the transport measurements including the zero temperature critical magnetic field (B$_0$), zero temperature critical current density (J$_0$), BCS superconducting gap ($\Delta \approx 1.76 \text{k}_\text{B} \text{T}_\text{C}$, where k$_B$ is the Boltzmann constant) and the superconducting coherence length ($\xi_0 = \sqrt{\frac{\Phi_0}{2 \pi B_0}}$, where $\Phi_0$ is the flux quantum).

The superconducting transition increases from 8.3 K for the as-grown sample to 9.2 K (value for bulk Nb) after UHV annealing at 300 $^{\circ}$C. On the other hand, annealing at higher temperatures such as 500 $^{\circ}$C degrades the superconducting properties of the films as evidenced by a significant drop in T$_\text{C}$ to 2.3 K. Similarly, J$_\text{C}$ increases after annealing at 300 $^{\circ}$C and significantly decreases at 500 $^{\circ}$C. The behavior can be attributed to the increased average grain size in the films (see  Supplementary Material, Fig. S3). However, the influence of larger grain size is counteracted by large Ge content at 500 $^{\circ}$C leading to degraded superconducting properties. Our reported J$_0$ values are based on fitting the experimental data to the Ginzburg-Landau (GL) theory for depairing critical current: J$_\text{C}$(t) =J$_0$(1-t$^2$)$^{\alpha}$ (1+t$^2$)$^{1/2}$ , where t = T/T$_\text{C}$ (See Supplementary Material, Fig. S4). We note that the temperature dependence of J$_\text{C}$ deviates from the conventional GL theory where $\alpha$  = 1.5 for samples that were as-grown or annealed at 500 $^{\circ}$C ($\alpha$ $\approx$ 1).~\cite{ilin_critical_2005, PhysRev.140.A1568} This deviation may be due to the disorder in the two films. In the former case, the disorder stems from the nanocrysallinity of the cryogenically-grown Nb samples. In the latter case, the large Ge content leads to atomic scale disorder within the Nb-Ge grains (see Supplementary Material, Fig. S2).

\begin{table}
\caption{\label{tab:table2}Summary of the superconducting characteristics measured for three Nb/Ge samples with different thermal processing conditions from as-grown to annealing at 300 $^{\circ}$C and 500 $^{\circ}$C. The parameters include superconducting transition temperature (T$_\text{C}$), the out-of-plane critical magnetic field at zero temperature (B$_0$), the critical current density at zero temperature (J$_0$), the BCS superconducting gap ($\Delta \approx 1.76 \text{k}_\text{B} \text{T}_\text{C}$), and the superconducting coherence length ($\xi_0$).}
\begin{ruledtabular}
\begin{tabular}{cccccc}
Anneal Temperature&T$_\text{C}$ (K)&B$_0$ (T)&J$_0$ (A/cm$^2$)&$\Delta(meV)$&$\xi_0$(nm)\\
\hline
As-grown & 8.3& 5.3 & 8.62x10$^{6}$&1.26&11.1 \\
300 $^{\circ}$C & 9.2& 2.2 & 6.07x10$^{7}$&1.40&17.20 \\
500 $^{\circ}$C & 2.3& 0.1 & 2.09x10$^{6}$&0.35&70.16 \\
\end{tabular}
\end{ruledtabular}
\end{table}

Unlike T$_\text{C}$ and J$_\text{C}$(T), B$_\text{C}$(T) and B$_0$ values significantly declined with annealing temperature. The as-grown sample has the highest B$_0$ of 5.3 T. Annealing the Nb/Ge samples to 300 $^{\circ}$C and 500 $^{\circ}$C decreased B$_0$ to 2.2 T and 0.1 T, respectively. Despite the significant change in its absolute values, the B$_\text{C}$ for our Nb/Ge samples maintained a linear temperature dependence irrespective of the annealing conditions. B$_0$ and $\xi_0$ were estimated by fitting the B$_\text{C}$(T) data for each sample to the linear Ginzburg–Landau (GL) relationship B$_\text{C}$(t) =B$_0$(1-t) , where t = T/T$_\text{C}$. Linear B$_\text{C}$(T) has been previously reported in two-dimensional disordered superconducting films of Ge:Ga and InO.~\cite{sardashti_tuning_2021,sacepe_localization_2011} This picture, is once again consistent with a disorder in the Nb/Ge heterostructures due to either nanocrystalline Nb structures (as-grown cryogenically) or solid-state Nb-Ge alloying at temperatures as low as 300 $^{\circ}$C.


In summary, we demonstrated chemically abrupt and atomically sharp Nb/Ge interfaces by UHV evaporation of Nb on Ge substrates held at temperatures as low as 100 K. Cryogenic growth also minimizes the physical damage to the interface. As-grown Nb/Ge interfaces show a relatively low band bending (121 meV) that can benefit future hybrid S-Sm devices. Annealing the samples to temperatures above 575 $^{\circ}$C led to even smaller interface band bending at the cost of significant Ge incorporation (up to 65 at.\%) into the Nb thin films. Additionally, annealing at such high temperatures led to the degradation of superconducting properties in the Nb films including critical temperature, current, and magnetic field.  Our results show a low thermal budget for realizing low-barrier Nb-Ge interfaces making a low-temperature or a cryogenic Nb growth approach essential in fabricating future hybrid Nb-Ge-Nb devices.\\


The \textbf{supplementary material} provides more information on the fs-UPS, XPS, and AFM measurements. Details on the microfabrication of the Nb microwires are also included. Additional data includes AFM images of Nb surfaces as a function of anneal temperature, cross-sectional STEM of a Nb/Ge sample annealed to 650 $^{\circ}$C, temperature-dependence of Nb work function, and the fittings for critical current vs temperature.

\begin{acknowledgments}
This work is supported by the National Science Foundation ( award no. 2137776) and the US Department of Energy (Award no. DE-SC0023595). Fabrication of niobium microwires was conducted as part of a user project at the Center for Nanophase Materials Sciences (CNMS), which is a US Department of Energy, Office of Science User Facility at Oak Ridge National Laboratory. RBS measurements were done at the Laboratory of Surface Modifications at Rutgers University. B.L. and K.S. thank Kelliann Koehler and the Clemson Electron Microscopy Facility for their assistance conducting the XPS measurements. M.K. was partly supported by Louis Beecherl, Jr. Endowed fund. 
\end{acknowledgments}

\bibliography{NbGe_refs.bib}

\begin{thebibliography}{25}%
\makeatletter
\providecommand \@ifxundefined [1]{%
 \@ifx{#1\undefined}
}%
\providecommand \@ifnum [1]{%
 \ifnum #1\expandafter \@firstoftwo
 \else \expandafter \@secondoftwo
 \fi
}%
\providecommand \@ifx [1]{%
 \ifx #1\expandafter \@firstoftwo
 \else \expandafter \@secondoftwo
 \fi
}%
\providecommand \natexlab [1]{#1}%
\providecommand \enquote  [1]{``#1''}%
\providecommand \bibnamefont  [1]{#1}%
\providecommand \bibfnamefont [1]{#1}%
\providecommand \citenamefont [1]{#1}%
\providecommand \href@noop [0]{\@secondoftwo}%
\providecommand \href [0]{\begingroup \@sanitize@url \@href}%
\providecommand \@href[1]{\@@startlink{#1}\@@href}%
\providecommand \@@href[1]{\endgroup#1\@@endlink}%
\providecommand \@sanitize@url [0]{\catcode `\\12\catcode `\$12\catcode `\&12\catcode `\#12\catcode `\^12\catcode `\_12\catcode `\%12\relax}%
\providecommand \@@startlink[1]{}%
\providecommand \@@endlink[0]{}%
\providecommand \url  [0]{\begingroup\@sanitize@url \@url }%
\providecommand \@url [1]{\endgroup\@href {#1}{\urlprefix }}%
\providecommand \urlprefix  [0]{URL }%
\providecommand \Eprint [0]{\href }%
\providecommand \doibase [0]{http://dx.doi.org/}%
\providecommand \selectlanguage [0]{\@gobble}%
\providecommand \bibinfo  [0]{\@secondoftwo}%
\providecommand \bibfield  [0]{\@secondoftwo}%
\providecommand \translation [1]{[#1]}%
\providecommand \BibitemOpen [0]{}%
\providecommand \bibitemStop [0]{}%
\providecommand \bibitemNoStop [0]{.\EOS\space}%
\providecommand \EOS [0]{\spacefactor3000\relax}%
\providecommand \BibitemShut  [1]{\csname bibitem#1\endcsname}%
\let\auto@bib@innerbib\@empty
\bibitem [{\citenamefont {Sardashti}\ \emph {et~al.}(2020)\citenamefont {Sardashti}, \citenamefont {Dartiailh}, \citenamefont {Yuan}, \citenamefont {Hart}, \citenamefont {Gumann},\ and\ \citenamefont {Shabani}}]{sardashti_voltage-tunable_2020}%
  \BibitemOpen
  \bibfield  {author} {\bibinfo {author} {\bibfnamefont {K.}~\bibnamefont {Sardashti}}, \bibinfo {author} {\bibfnamefont {M.~C.}\ \bibnamefont {Dartiailh}}, \bibinfo {author} {\bibfnamefont {J.}~\bibnamefont {Yuan}}, \bibinfo {author} {\bibfnamefont {S.}~\bibnamefont {Hart}}, \bibinfo {author} {\bibfnamefont {P.}~\bibnamefont {Gumann}}, \ and\ \bibinfo {author} {\bibfnamefont {J.}~\bibnamefont {Shabani}},\ }\bibfield  {title} {{\selectlanguage {en}\enquote {\bibinfo {title} {Voltage-{Tunable} {Superconducting} {Resonators}: {A} {Platform} for {Random} {Access} {Quantum} {Memory}},}\ }}\href {\doibase 10.1109/TQE.2020.3034553} {\bibfield  {journal} {\bibinfo  {journal} {IEEE Transactions on Quantum Engineering}\ }\textbf {\bibinfo {volume} {1}},\ \bibinfo {pages} {1--7} (\bibinfo {year} {2020})}\BibitemShut {NoStop}%
\bibitem [{\citenamefont {Casparis}\ \emph {et~al.}(2019)\citenamefont {Casparis}, \citenamefont {Pearson}, \citenamefont {Kringhøj}, \citenamefont {Larsen}, \citenamefont {Kuemmeth}, \citenamefont {Nygård}, \citenamefont {Krogstrup}, \citenamefont {Petersson},\ and\ \citenamefont {Marcus}}]{casparis_voltage-controlled_2019}%
  \BibitemOpen
  \bibfield  {author} {\bibinfo {author} {\bibfnamefont {L.}~\bibnamefont {Casparis}}, \bibinfo {author} {\bibfnamefont {N.~J.}\ \bibnamefont {Pearson}}, \bibinfo {author} {\bibfnamefont {A.}~\bibnamefont {Kringhøj}}, \bibinfo {author} {\bibfnamefont {T.~W.}\ \bibnamefont {Larsen}}, \bibinfo {author} {\bibfnamefont {F.}~\bibnamefont {Kuemmeth}}, \bibinfo {author} {\bibfnamefont {J.}~\bibnamefont {Nygård}}, \bibinfo {author} {\bibfnamefont {P.}~\bibnamefont {Krogstrup}}, \bibinfo {author} {\bibfnamefont {K.~D.}\ \bibnamefont {Petersson}}, \ and\ \bibinfo {author} {\bibfnamefont {C.~M.}\ \bibnamefont {Marcus}},\ }\bibfield  {title} {\enquote {\bibinfo {title} {Voltage-controlled superconducting quantum bus},}\ }\href {\doibase 10.1103/PhysRevB.99.085434} {\bibfield  {journal} {\bibinfo  {journal} {Physical Review B}\ }\textbf {\bibinfo {volume} {99}},\ \bibinfo {pages} {085434} (\bibinfo {year} {2019})},\ \bibinfo {note} {publisher: American Physical Society}\BibitemShut {NoStop}%
\bibitem [{\citenamefont {Strickland}\ \emph {et~al.}(2023)\citenamefont {Strickland}, \citenamefont {Elfeky}, \citenamefont {Yuan}, \citenamefont {Schiela}, \citenamefont {Yu}, \citenamefont {Langone}, \citenamefont {Vavilov}, \citenamefont {Manucharyan},\ and\ \citenamefont {Shabani}}]{strickland_superconducting_2023}%
  \BibitemOpen
  \bibfield  {author} {\bibinfo {author} {\bibfnamefont {W.~M.}\ \bibnamefont {Strickland}}, \bibinfo {author} {\bibfnamefont {B.~H.}\ \bibnamefont {Elfeky}}, \bibinfo {author} {\bibfnamefont {J.~O.}\ \bibnamefont {Yuan}}, \bibinfo {author} {\bibfnamefont {W.~F.}\ \bibnamefont {Schiela}}, \bibinfo {author} {\bibfnamefont {P.}~\bibnamefont {Yu}}, \bibinfo {author} {\bibfnamefont {D.}~\bibnamefont {Langone}}, \bibinfo {author} {\bibfnamefont {M.~G.}\ \bibnamefont {Vavilov}}, \bibinfo {author} {\bibfnamefont {V.~E.}\ \bibnamefont {Manucharyan}}, \ and\ \bibinfo {author} {\bibfnamefont {J.}~\bibnamefont {Shabani}},\ }\bibfield  {title} {{\selectlanguage {en}\enquote {\bibinfo {title} {Superconducting {Resonators} with {Voltage}-{Controlled} {Frequency} and {Nonlinearity}},}\ }}\href {\doibase 10.1103/PhysRevApplied.19.034021} {\bibfield  {journal} {\bibinfo  {journal} {Physical Review Applied}\ }\textbf {\bibinfo {volume} {19}},\ \bibinfo {pages} {034021} (\bibinfo {year} {2023})}\BibitemShut {NoStop}%
\bibitem [{\citenamefont {Larsen}\ \emph {et~al.}(2015)\citenamefont {Larsen}, \citenamefont {Petersson}, \citenamefont {Kuemmeth}, \citenamefont {Jespersen}, \citenamefont {Krogstrup}, \citenamefont {Nygård},\ and\ \citenamefont {Marcus}}]{larsen_semiconductor-nanowire-based_2015}%
  \BibitemOpen
  \bibfield  {author} {\bibinfo {author} {\bibfnamefont {T.}~\bibnamefont {Larsen}}, \bibinfo {author} {\bibfnamefont {K.}~\bibnamefont {Petersson}}, \bibinfo {author} {\bibfnamefont {F.}~\bibnamefont {Kuemmeth}}, \bibinfo {author} {\bibfnamefont {T.}~\bibnamefont {Jespersen}}, \bibinfo {author} {\bibfnamefont {P.}~\bibnamefont {Krogstrup}}, \bibinfo {author} {\bibfnamefont {J.}~\bibnamefont {Nygård}}, \ and\ \bibinfo {author} {\bibfnamefont {C.}~\bibnamefont {Marcus}},\ }\bibfield  {title} {\enquote {\bibinfo {title} {Semiconductor-{Nanowire}-{Based} {Superconducting} {Qubit}},}\ }\href {\doibase 10.1103/PhysRevLett.115.127001} {\bibfield  {journal} {\bibinfo  {journal} {Physical Review Letters}\ }\textbf {\bibinfo {volume} {115}},\ \bibinfo {pages} {127001} (\bibinfo {year} {2015})},\ \bibinfo {note} {publisher: American Physical Society}\BibitemShut {NoStop}%
\bibitem [{\citenamefont {Casparis}\ \emph {et~al.}(2018)\citenamefont {Casparis}, \citenamefont {Connolly}, \citenamefont {Kjaergaard}, \citenamefont {Pearson}, \citenamefont {Kringhøj}, \citenamefont {Larsen}, \citenamefont {Kuemmeth}, \citenamefont {Wang}, \citenamefont {Thomas}, \citenamefont {Gronin}, \citenamefont {Gardner}, \citenamefont {Manfra}, \citenamefont {Marcus},\ and\ \citenamefont {Petersson}}]{casparis_superconducting_2018}%
  \BibitemOpen
  \bibfield  {author} {\bibinfo {author} {\bibfnamefont {L.}~\bibnamefont {Casparis}}, \bibinfo {author} {\bibfnamefont {M.~R.}\ \bibnamefont {Connolly}}, \bibinfo {author} {\bibfnamefont {M.}~\bibnamefont {Kjaergaard}}, \bibinfo {author} {\bibfnamefont {N.~J.}\ \bibnamefont {Pearson}}, \bibinfo {author} {\bibfnamefont {A.}~\bibnamefont {Kringhøj}}, \bibinfo {author} {\bibfnamefont {T.~W.}\ \bibnamefont {Larsen}}, \bibinfo {author} {\bibfnamefont {F.}~\bibnamefont {Kuemmeth}}, \bibinfo {author} {\bibfnamefont {T.}~\bibnamefont {Wang}}, \bibinfo {author} {\bibfnamefont {C.}~\bibnamefont {Thomas}}, \bibinfo {author} {\bibfnamefont {S.}~\bibnamefont {Gronin}}, \bibinfo {author} {\bibfnamefont {G.~C.}\ \bibnamefont {Gardner}}, \bibinfo {author} {\bibfnamefont {M.~J.}\ \bibnamefont {Manfra}}, \bibinfo {author} {\bibfnamefont {C.~M.}\ \bibnamefont {Marcus}}, \ and\ \bibinfo {author} {\bibfnamefont {K.~D.}\ \bibnamefont {Petersson}},\ }\bibfield  {title} {{\selectlanguage {en}\enquote {\bibinfo {title}
  {Superconducting gatemon qubit based on a proximitized two-dimensional electron gas},}\ }}\href {\doibase 10.1038/s41565-018-0207-y} {\bibfield  {journal} {\bibinfo  {journal} {Nature Nanotechnology}\ }\textbf {\bibinfo {volume} {13}},\ \bibinfo {pages} {915--919} (\bibinfo {year} {2018})},\ \bibinfo {note} {publisher: Nature Publishing Group}\BibitemShut {NoStop}%
\bibitem [{\citenamefont {Klapwijk}(2004)}]{klapwijk_proximity_2004}%
  \BibitemOpen
  \bibfield  {author} {\bibinfo {author} {\bibfnamefont {T.}~\bibnamefont {Klapwijk}},\ }\bibfield  {title} {\enquote {\bibinfo {title} {Proximity {Effect} {From} an {Andreev} {Perspective}},}\ }\href {\doibase 10.1007/s10948-004-0773-0} {\bibfield  {journal} {\bibinfo  {journal} {Journal of Superconductivity}\ }\textbf {\bibinfo {volume} {17}},\ \bibinfo {pages} {593--611} (\bibinfo {year} {2004})}\BibitemShut {NoStop}%
\bibitem [{\citenamefont {Shabani}\ \emph {et~al.}(2016)\citenamefont {Shabani}, \citenamefont {Kjaergaard}, \citenamefont {Suominen}, \citenamefont {Kim}, \citenamefont {Nichele}, \citenamefont {Pakrouski}, \citenamefont {Stankevic}, \citenamefont {Lutchyn}, \citenamefont {Krogstrup}, \citenamefont {Feidenhans'l}, \citenamefont {Kraemer}, \citenamefont {Nayak}, \citenamefont {Troyer}, \citenamefont {Marcus},\ and\ \citenamefont {Palmstrøm}}]{shabani_two-dimensional_2016}%
  \BibitemOpen
  \bibfield  {author} {\bibinfo {author} {\bibfnamefont {J.}~\bibnamefont {Shabani}}, \bibinfo {author} {\bibfnamefont {M.}~\bibnamefont {Kjaergaard}}, \bibinfo {author} {\bibfnamefont {H.~J.}\ \bibnamefont {Suominen}}, \bibinfo {author} {\bibfnamefont {Y.}~\bibnamefont {Kim}}, \bibinfo {author} {\bibfnamefont {F.}~\bibnamefont {Nichele}}, \bibinfo {author} {\bibfnamefont {K.}~\bibnamefont {Pakrouski}}, \bibinfo {author} {\bibfnamefont {T.}~\bibnamefont {Stankevic}}, \bibinfo {author} {\bibfnamefont {R.~M.}\ \bibnamefont {Lutchyn}}, \bibinfo {author} {\bibfnamefont {P.}~\bibnamefont {Krogstrup}}, \bibinfo {author} {\bibfnamefont {R.}~\bibnamefont {Feidenhans'l}}, \bibinfo {author} {\bibfnamefont {S.}~\bibnamefont {Kraemer}}, \bibinfo {author} {\bibfnamefont {C.}~\bibnamefont {Nayak}}, \bibinfo {author} {\bibfnamefont {M.}~\bibnamefont {Troyer}}, \bibinfo {author} {\bibfnamefont {C.~M.}\ \bibnamefont {Marcus}}, \ and\ \bibinfo {author} {\bibfnamefont {C.~J.}\ \bibnamefont {Palmstrøm}},\ }\bibfield  {title}
  {\enquote {\bibinfo {title} {Two-dimensional epitaxial superconductor-semiconductor heterostructures: {A} platform for topological superconducting networks},}\ }\href {\doibase 10.1103/PhysRevB.93.155402} {\bibfield  {journal} {\bibinfo  {journal} {Physical Review B}\ }\textbf {\bibinfo {volume} {93}},\ \bibinfo {pages} {155402} (\bibinfo {year} {2016})},\ \bibinfo {note} {arXiv:1511.01127 [cond-mat]}\BibitemShut {NoStop}%
\bibitem [{\citenamefont {Wickramasinghe}\ \emph {et~al.}(2018)\citenamefont {Wickramasinghe}, \citenamefont {Mayer}, \citenamefont {Yuan}, \citenamefont {Nguyen}, \citenamefont {Jiao}, \citenamefont {Manucharyan},\ and\ \citenamefont {Shabani}}]{wickramasinghe_transport_2018}%
  \BibitemOpen
  \bibfield  {author} {\bibinfo {author} {\bibfnamefont {K.}~\bibnamefont {Wickramasinghe}}, \bibinfo {author} {\bibfnamefont {W.}~\bibnamefont {Mayer}}, \bibinfo {author} {\bibfnamefont {J.}~\bibnamefont {Yuan}}, \bibinfo {author} {\bibfnamefont {T.}~\bibnamefont {Nguyen}}, \bibinfo {author} {\bibfnamefont {L.}~\bibnamefont {Jiao}}, \bibinfo {author} {\bibfnamefont {V.}~\bibnamefont {Manucharyan}}, \ and\ \bibinfo {author} {\bibfnamefont {J.}~\bibnamefont {Shabani}},\ }\bibfield  {title} {\enquote {\bibinfo {title} {Transport {Properties} of {Near} {Surface} {InAs} {Two}-dimensional {Heterostructures}},}\ }\href {\doibase 10.1063/1.5050413} {\bibfield  {journal} {\bibinfo  {journal} {Applied Physics Letters}\ }\textbf {\bibinfo {volume} {113}},\ \bibinfo {pages} {262104} (\bibinfo {year} {2018})},\ \bibinfo {note} {arXiv:1802.09569 [cond-mat]}\BibitemShut {NoStop}%
\bibitem [{\citenamefont {Kringhøj}\ \emph {et~al.}(2021)\citenamefont {Kringhøj}, \citenamefont {Larsen}, \citenamefont {Erlandsson}, \citenamefont {Uilhoorn}, \citenamefont {Kroll}, \citenamefont {Hesselberg}, \citenamefont {McNeil}, \citenamefont {Krogstrup}, \citenamefont {Casparis}, \citenamefont {Marcus},\ and\ \citenamefont {Petersson}}]{kringhoj_magnetic-field-compatible_2021}%
  \BibitemOpen
  \bibfield  {author} {\bibinfo {author} {\bibfnamefont {A.}~\bibnamefont {Kringhøj}}, \bibinfo {author} {\bibfnamefont {T.~W.}\ \bibnamefont {Larsen}}, \bibinfo {author} {\bibfnamefont {O.}~\bibnamefont {Erlandsson}}, \bibinfo {author} {\bibfnamefont {W.}~\bibnamefont {Uilhoorn}}, \bibinfo {author} {\bibfnamefont {J.}~\bibnamefont {Kroll}}, \bibinfo {author} {\bibfnamefont {M.}~\bibnamefont {Hesselberg}}, \bibinfo {author} {\bibfnamefont {R.}~\bibnamefont {McNeil}}, \bibinfo {author} {\bibfnamefont {P.}~\bibnamefont {Krogstrup}}, \bibinfo {author} {\bibfnamefont {L.}~\bibnamefont {Casparis}}, \bibinfo {author} {\bibfnamefont {C.}~\bibnamefont {Marcus}}, \ and\ \bibinfo {author} {\bibfnamefont {K.}~\bibnamefont {Petersson}},\ }\bibfield  {title} {\enquote {\bibinfo {title} {Magnetic-{Field}-{Compatible} {Superconducting} {Transmon} {Qubit}},}\ }\href {\doibase 10.1103/PhysRevApplied.15.054001} {\bibfield  {journal} {\bibinfo  {journal} {Physical Review Applied}\ }\textbf {\bibinfo {volume} {15}},\ \bibinfo
  {pages} {054001} (\bibinfo {year} {2021})},\ \bibinfo {note} {publisher: American Physical Society}\BibitemShut {NoStop}%
\bibitem [{\citenamefont {O’Connell~Yuan}\ \emph {et~al.}(2021)\citenamefont {O’Connell~Yuan}, \citenamefont {Wickramasinghe}, \citenamefont {Strickland}, \citenamefont {Dartiailh}, \citenamefont {Sardashti}, \citenamefont {Hatefipour},\ and\ \citenamefont {Shabani}}]{oconnell_yuan_epitaxial_2021}%
  \BibitemOpen
  \bibfield  {author} {\bibinfo {author} {\bibfnamefont {J.}~\bibnamefont {O’Connell~Yuan}}, \bibinfo {author} {\bibfnamefont {K.~S.}\ \bibnamefont {Wickramasinghe}}, \bibinfo {author} {\bibfnamefont {W.~M.}\ \bibnamefont {Strickland}}, \bibinfo {author} {\bibfnamefont {M.~C.}\ \bibnamefont {Dartiailh}}, \bibinfo {author} {\bibfnamefont {K.}~\bibnamefont {Sardashti}}, \bibinfo {author} {\bibfnamefont {M.}~\bibnamefont {Hatefipour}}, \ and\ \bibinfo {author} {\bibfnamefont {J.}~\bibnamefont {Shabani}},\ }\bibfield  {title} {{\selectlanguage {en}\enquote {\bibinfo {title} {Epitaxial superconductor-semiconductor two-dimensional systems for superconducting quantum circuits},}\ }}\href {\doibase 10.1116/6.0000918} {\bibfield  {journal} {\bibinfo  {journal} {Journal of Vacuum Science \& Technology A: Vacuum, Surfaces, and Films}\ }\textbf {\bibinfo {volume} {39}},\ \bibinfo {pages} {033407} (\bibinfo {year} {2021})}\BibitemShut {NoStop}%
\bibitem [{\citenamefont {Strickland}\ \emph {et~al.}(2024)\citenamefont {Strickland}, \citenamefont {Baker}, \citenamefont {Lee}, \citenamefont {Dindial}, \citenamefont {Elfeky}, \citenamefont {Strohbeen}, \citenamefont {Hatefipour}, \citenamefont {Yu}, \citenamefont {Levy}, \citenamefont {Issokson}, \citenamefont {Manucharyan},\ and\ \citenamefont {Shabani}}]{strickland_characterizing_2024}%
  \BibitemOpen
  \bibfield  {author} {\bibinfo {author} {\bibfnamefont {W.~M.}\ \bibnamefont {Strickland}}, \bibinfo {author} {\bibfnamefont {L.~J.}\ \bibnamefont {Baker}}, \bibinfo {author} {\bibfnamefont {J.}~\bibnamefont {Lee}}, \bibinfo {author} {\bibfnamefont {K.}~\bibnamefont {Dindial}}, \bibinfo {author} {\bibfnamefont {B.~H.}\ \bibnamefont {Elfeky}}, \bibinfo {author} {\bibfnamefont {P.~J.}\ \bibnamefont {Strohbeen}}, \bibinfo {author} {\bibfnamefont {M.}~\bibnamefont {Hatefipour}}, \bibinfo {author} {\bibfnamefont {P.}~\bibnamefont {Yu}}, \bibinfo {author} {\bibfnamefont {I.}~\bibnamefont {Levy}}, \bibinfo {author} {\bibfnamefont {J.}~\bibnamefont {Issokson}}, \bibinfo {author} {\bibfnamefont {V.~E.}\ \bibnamefont {Manucharyan}}, \ and\ \bibinfo {author} {\bibfnamefont {J.}~\bibnamefont {Shabani}},\ }\bibfield  {title} {{\selectlanguage {en}\enquote {\bibinfo {title} {Characterizing losses in {InAs} two-dimensional electron gas-based gatemon qubits},}\ }}\href {\doibase 10.1103/PhysRevResearch.6.023094} {\bibfield
  {journal} {\bibinfo  {journal} {Physical Review Research}\ }\textbf {\bibinfo {volume} {6}},\ \bibinfo {pages} {023094} (\bibinfo {year} {2024})}\BibitemShut {NoStop}%
\bibitem [{\citenamefont {Vigneau}\ \emph {et~al.}(2019)\citenamefont {Vigneau}, \citenamefont {Mizokuchi}, \citenamefont {Zanuz}, \citenamefont {Huang}, \citenamefont {Tan}, \citenamefont {Maurand}, \citenamefont {Frolov}, \citenamefont {Sammak}, \citenamefont {Scappucci}, \citenamefont {Lefloch},\ and\ \citenamefont {De~Franceschi}}]{vigneau_germanium_2019}%
  \BibitemOpen
  \bibfield  {author} {\bibinfo {author} {\bibfnamefont {F.}~\bibnamefont {Vigneau}}, \bibinfo {author} {\bibfnamefont {R.}~\bibnamefont {Mizokuchi}}, \bibinfo {author} {\bibfnamefont {D.~C.}\ \bibnamefont {Zanuz}}, \bibinfo {author} {\bibfnamefont {X.}~\bibnamefont {Huang}}, \bibinfo {author} {\bibfnamefont {S.}~\bibnamefont {Tan}}, \bibinfo {author} {\bibfnamefont {R.}~\bibnamefont {Maurand}}, \bibinfo {author} {\bibfnamefont {S.}~\bibnamefont {Frolov}}, \bibinfo {author} {\bibfnamefont {A.}~\bibnamefont {Sammak}}, \bibinfo {author} {\bibfnamefont {G.}~\bibnamefont {Scappucci}}, \bibinfo {author} {\bibfnamefont {F.}~\bibnamefont {Lefloch}}, \ and\ \bibinfo {author} {\bibfnamefont {S.}~\bibnamefont {De~Franceschi}},\ }\bibfield  {title} {\enquote {\bibinfo {title} {Germanium {Quantum}-{Well} {Josephson} {Field}-{Effect} {Transistors} and {Interferometers}},}\ }\href {\doibase 10.1021/acs.nanolett.8b04275} {\bibfield  {journal} {\bibinfo  {journal} {Nano Letters}\ }\textbf {\bibinfo {volume} {19}},\ \bibinfo
  {pages} {1023--1027} (\bibinfo {year} {2019})},\ \bibinfo {note} {publisher: American Chemical Society}\BibitemShut {NoStop}%
\bibitem [{\citenamefont {Tosato}\ \emph {et~al.}(2023)\citenamefont {Tosato}, \citenamefont {Levajac}, \citenamefont {Wang}, \citenamefont {Boor}, \citenamefont {Borsoi}, \citenamefont {Botifoll}, \citenamefont {Borja}, \citenamefont {Martí-Sánchez}, \citenamefont {Arbiol}, \citenamefont {Sammak}, \citenamefont {Veldhorst},\ and\ \citenamefont {Scappucci}}]{tosato_hard_2023}%
  \BibitemOpen
  \bibfield  {author} {\bibinfo {author} {\bibfnamefont {A.}~\bibnamefont {Tosato}}, \bibinfo {author} {\bibfnamefont {V.}~\bibnamefont {Levajac}}, \bibinfo {author} {\bibfnamefont {J.-Y.}\ \bibnamefont {Wang}}, \bibinfo {author} {\bibfnamefont {C.~J.}\ \bibnamefont {Boor}}, \bibinfo {author} {\bibfnamefont {F.}~\bibnamefont {Borsoi}}, \bibinfo {author} {\bibfnamefont {M.}~\bibnamefont {Botifoll}}, \bibinfo {author} {\bibfnamefont {C.~N.}\ \bibnamefont {Borja}}, \bibinfo {author} {\bibfnamefont {S.}~\bibnamefont {Martí-Sánchez}}, \bibinfo {author} {\bibfnamefont {J.}~\bibnamefont {Arbiol}}, \bibinfo {author} {\bibfnamefont {A.}~\bibnamefont {Sammak}}, \bibinfo {author} {\bibfnamefont {M.}~\bibnamefont {Veldhorst}}, \ and\ \bibinfo {author} {\bibfnamefont {G.}~\bibnamefont {Scappucci}},\ }\bibfield  {title} {{\selectlanguage {en}\enquote {\bibinfo {title} {Hard superconducting gap in germanium},}\ }}\href {\doibase 10.1038/s43246-023-00351-w} {\bibfield  {journal} {\bibinfo  {journal} {Communications
  Materials}\ }\textbf {\bibinfo {volume} {4}},\ \bibinfo {pages} {1--9} (\bibinfo {year} {2023})},\ \bibinfo {note} {publisher: Nature Publishing Group}\BibitemShut {NoStop}%
\bibitem [{\citenamefont {Aggarwal}\ \emph {et~al.}(2021)\citenamefont {Aggarwal}, \citenamefont {Hofmann}, \citenamefont {Jirovec}, \citenamefont {Prieto}, \citenamefont {Sammak}, \citenamefont {Botifoll}, \citenamefont {Martí-Sánchez}, \citenamefont {Veldhorst}, \citenamefont {Arbiol}, \citenamefont {Scappucci}, \citenamefont {Danon},\ and\ \citenamefont {Katsaros}}]{aggarwal_enhancement_2021}%
  \BibitemOpen
  \bibfield  {author} {\bibinfo {author} {\bibfnamefont {K.}~\bibnamefont {Aggarwal}}, \bibinfo {author} {\bibfnamefont {A.}~\bibnamefont {Hofmann}}, \bibinfo {author} {\bibfnamefont {D.}~\bibnamefont {Jirovec}}, \bibinfo {author} {\bibfnamefont {I.}~\bibnamefont {Prieto}}, \bibinfo {author} {\bibfnamefont {A.}~\bibnamefont {Sammak}}, \bibinfo {author} {\bibfnamefont {M.}~\bibnamefont {Botifoll}}, \bibinfo {author} {\bibfnamefont {S.}~\bibnamefont {Martí-Sánchez}}, \bibinfo {author} {\bibfnamefont {M.}~\bibnamefont {Veldhorst}}, \bibinfo {author} {\bibfnamefont {J.}~\bibnamefont {Arbiol}}, \bibinfo {author} {\bibfnamefont {G.}~\bibnamefont {Scappucci}}, \bibinfo {author} {\bibfnamefont {J.}~\bibnamefont {Danon}}, \ and\ \bibinfo {author} {\bibfnamefont {G.}~\bibnamefont {Katsaros}},\ }\bibfield  {title} {\enquote {\bibinfo {title} {Enhancement of proximity-induced superconductivity in a planar {Ge} hole gas},}\ }\href {\doibase 10.1103/PhysRevResearch.3.L022005} {\bibfield  {journal} {\bibinfo  {journal}
  {Physical Review Research}\ }\textbf {\bibinfo {volume} {3}},\ \bibinfo {pages} {L022005} (\bibinfo {year} {2021})},\ \bibinfo {note} {publisher: American Physical Society}\BibitemShut {NoStop}%
\bibitem [{\citenamefont {Iijima}, \citenamefont {Igarashi},\ and\ \citenamefont {Hirano}(1979)}]{iijima_reaction_1979}%
  \BibitemOpen
  \bibfield  {author} {\bibinfo {author} {\bibfnamefont {Y.}~\bibnamefont {Iijima}}, \bibinfo {author} {\bibfnamefont {T.}~\bibnamefont {Igarashi}}, \ and\ \bibinfo {author} {\bibfnamefont {K.~I.}\ \bibnamefont {Hirano}},\ }\bibfield  {title} {{\selectlanguage {en}\enquote {\bibinfo {title} {Reaction diffusion in the {Nb}-{Ge} system},}\ }}\href {\doibase 10.1007/BF00589842} {\bibfield  {journal} {\bibinfo  {journal} {Journal of Materials Science}\ }\textbf {\bibinfo {volume} {14}},\ \bibinfo {pages} {474--479} (\bibinfo {year} {1979})}\BibitemShut {NoStop}%
\bibitem [{\citenamefont {Lim}\ \emph {et~al.}(2005)\citenamefont {Lim}, \citenamefont {Haight}, \citenamefont {Copel},\ and\ \citenamefont {Cartier}}]{lim_oxygen_2005}%
  \BibitemOpen
  \bibfield  {author} {\bibinfo {author} {\bibfnamefont {D.}~\bibnamefont {Lim}}, \bibinfo {author} {\bibfnamefont {R.}~\bibnamefont {Haight}}, \bibinfo {author} {\bibfnamefont {M.}~\bibnamefont {Copel}}, \ and\ \bibinfo {author} {\bibfnamefont {E.}~\bibnamefont {Cartier}},\ }\bibfield  {title} {\enquote {\bibinfo {title} {Oxygen defects and {Fermi} level location in metal-hafnium oxide-silicon structures},}\ }\href {\doibase 10.1063/1.2011791} {\bibfield  {journal} {\bibinfo  {journal} {Applied Physics Letters}\ }\textbf {\bibinfo {volume} {87}},\ \bibinfo {pages} {072902} (\bibinfo {year} {2005})}\BibitemShut {NoStop}%
\bibitem [{\citenamefont {Lim}\ and\ \citenamefont {Haight}(2005{\natexlab{a}})}]{lim_situ_2005}%
  \BibitemOpen
  \bibfield  {author} {\bibinfo {author} {\bibfnamefont {D.}~\bibnamefont {Lim}}\ and\ \bibinfo {author} {\bibfnamefont {R.}~\bibnamefont {Haight}},\ }\bibfield  {title} {\enquote {\bibinfo {title} {In situ photovoltage measurements using femtosecond pump-probe photoelectron spectroscopy and its application to metal–{HfO2}–{Si} structures},}\ }\href {\doibase 10.1116/1.2083909} {\bibfield  {journal} {\bibinfo  {journal} {Journal of Vacuum Science \& Technology A}\ }\textbf {\bibinfo {volume} {23}},\ \bibinfo {pages} {1698--1705} (\bibinfo {year} {2005}{\natexlab{a}})}\BibitemShut {NoStop}%
\bibitem [{\citenamefont {Lim}\ and\ \citenamefont {Haight}(2005{\natexlab{b}})}]{lim_temperature_2005}%
  \BibitemOpen
  \bibfield  {author} {\bibinfo {author} {\bibfnamefont {D.}~\bibnamefont {Lim}}\ and\ \bibinfo {author} {\bibfnamefont {R.}~\bibnamefont {Haight}},\ }\bibfield  {title} {\enquote {\bibinfo {title} {Temperature dependent defect formation and charging in hafnium oxides and silicates},}\ }\href {\doibase 10.1116/1.1850105} {\bibfield  {journal} {\bibinfo  {journal} {Journal of Vacuum Science \& Technology B: Microelectronics and Nanometer Structures Processing, Measurement, and Phenomena}\ }\textbf {\bibinfo {volume} {23}},\ \bibinfo {pages} {201--205} (\bibinfo {year} {2005}{\natexlab{b}})}\BibitemShut {NoStop}%
\bibitem [{\citenamefont {Sasaki}\ and\ \citenamefont {Baba}(1985)}]{sasaki_chemical-state_1985}%
  \BibitemOpen
  \bibfield  {author} {\bibinfo {author} {\bibfnamefont {T.~A.}\ \bibnamefont {Sasaki}}\ and\ \bibinfo {author} {\bibfnamefont {Y.}~\bibnamefont {Baba}},\ }\bibfield  {title} {\enquote {\bibinfo {title} {Chemical-state studies of {Zr} and {Nb} surfaces exposed to hydrogen ions},}\ }\href {\doibase 10.1103/PhysRevB.31.791} {\bibfield  {journal} {\bibinfo  {journal} {Physical Review B}\ }\textbf {\bibinfo {volume} {31}},\ \bibinfo {pages} {791--797} (\bibinfo {year} {1985})},\ \bibinfo {note} {publisher: American Physical Society}\BibitemShut {NoStop}%
\bibitem [{\citenamefont {Morgan}\ and\ \citenamefont {Van~Wazer}(1973)}]{morgan_binding_1973}%
  \BibitemOpen
  \bibfield  {author} {\bibinfo {author} {\bibfnamefont {W.~E.}\ \bibnamefont {Morgan}}\ and\ \bibinfo {author} {\bibfnamefont {J.~R.}\ \bibnamefont {Van~Wazer}},\ }\bibfield  {title} {\enquote {\bibinfo {title} {Binding energy shifts in the x-ray photoelectron spectra of a series of related {Group} {IVa} compounds},}\ }\href {\doibase 10.1021/j100626a023} {\bibfield  {journal} {\bibinfo  {journal} {The Journal of Physical Chemistry}\ }\textbf {\bibinfo {volume} {77}},\ \bibinfo {pages} {964--969} (\bibinfo {year} {1973})},\ \bibinfo {note} {publisher: American Chemical Society}\BibitemShut {NoStop}%
\bibitem [{\citenamefont {Tanuma}, \citenamefont {Powell},\ and\ \citenamefont {Penn}(1994)}]{tanuma_calculations_1994}%
  \BibitemOpen
  \bibfield  {author} {\bibinfo {author} {\bibfnamefont {S.}~\bibnamefont {Tanuma}}, \bibinfo {author} {\bibfnamefont {C.~J.}\ \bibnamefont {Powell}}, \ and\ \bibinfo {author} {\bibfnamefont {D.~R.}\ \bibnamefont {Penn}},\ }\bibfield  {title} {{\selectlanguage {en}\enquote {\bibinfo {title} {Calculations of electron inelastic mean free paths. {V}. {Data} for 14 organic compounds over the 50–2000 {eV} range},}\ }}\href@noop {} {\bibfield  {journal} {\bibinfo  {journal} {Surface and Interface Analysis}\ }\textbf {\bibinfo {volume} {21}},\ \bibinfo {pages} {165--176} (\bibinfo {year} {1994})}\BibitemShut {NoStop}%
\bibitem [{\citenamefont {Il'in}\ \emph {et~al.}(2005)\citenamefont {Il'in}, \citenamefont {Siegel}, \citenamefont {Semenov}, \citenamefont {Engel},\ and\ \citenamefont {Hübers}}]{ilin_critical_2005}%
  \BibitemOpen
  \bibfield  {author} {\bibinfo {author} {\bibfnamefont {K.}~\bibnamefont {Il'in}}, \bibinfo {author} {\bibfnamefont {M.}~\bibnamefont {Siegel}}, \bibinfo {author} {\bibfnamefont {A.}~\bibnamefont {Semenov}}, \bibinfo {author} {\bibfnamefont {A.}~\bibnamefont {Engel}}, \ and\ \bibinfo {author} {\bibfnamefont {H.-W.}\ \bibnamefont {Hübers}},\ }\bibfield  {title} {{\selectlanguage {en}\enquote {\bibinfo {title} {Critical current of {Nb} and {NbN} thin-film structures: {The} cross-section dependence},}\ }}\href {\doibase 10.1002/pssc.200460811} {\bibfield  {journal} {\bibinfo  {journal} {physica status solidi (c)}\ }\textbf {\bibinfo {volume} {2}},\ \bibinfo {pages} {1680--1687} (\bibinfo {year} {2005})},\ \bibinfo {note} {\_eprint: https://onlinelibrary.wiley.com/doi/pdf/10.1002/pssc.200460811}\BibitemShut {NoStop}%
\bibitem [{\citenamefont {Mydosh}\ and\ \citenamefont {Meissner}(1965)}]{PhysRev.140.A1568}%
  \BibitemOpen
  \bibfield  {author} {\bibinfo {author} {\bibfnamefont {J.~A.}\ \bibnamefont {Mydosh}}\ and\ \bibinfo {author} {\bibfnamefont {H.}~\bibnamefont {Meissner}},\ }\bibfield  {title} {\enquote {\bibinfo {title} {Dependence of the critical currents in superconducting films on applied magnetic field and temperature},}\ }\href {\doibase 10.1103/PhysRev.140.A1568} {\bibfield  {journal} {\bibinfo  {journal} {Phys. Rev.}\ }\textbf {\bibinfo {volume} {140}},\ \bibinfo {pages} {A1568--A1580} (\bibinfo {year} {1965})}\BibitemShut {NoStop}%
\bibitem [{\citenamefont {Sardashti}\ \emph {et~al.}(2021)\citenamefont {Sardashti}, \citenamefont {Nguyen}, \citenamefont {Sarney}, \citenamefont {Leff}, \citenamefont {Hatefipour}, \citenamefont {Dartiailh}, \citenamefont {Yuan}, \citenamefont {Mayer},\ and\ \citenamefont {Shabani}}]{sardashti_tuning_2021}%
  \BibitemOpen
  \bibfield  {author} {\bibinfo {author} {\bibfnamefont {K.}~\bibnamefont {Sardashti}}, \bibinfo {author} {\bibfnamefont {T.~D.}\ \bibnamefont {Nguyen}}, \bibinfo {author} {\bibfnamefont {W.~L.}\ \bibnamefont {Sarney}}, \bibinfo {author} {\bibfnamefont {A.~C.}\ \bibnamefont {Leff}}, \bibinfo {author} {\bibfnamefont {M.}~\bibnamefont {Hatefipour}}, \bibinfo {author} {\bibfnamefont {M.~C.}\ \bibnamefont {Dartiailh}}, \bibinfo {author} {\bibfnamefont {J.}~\bibnamefont {Yuan}}, \bibinfo {author} {\bibfnamefont {W.}~\bibnamefont {Mayer}}, \ and\ \bibinfo {author} {\bibfnamefont {J.}~\bibnamefont {Shabani}},\ }\bibfield  {title} {\enquote {\bibinfo {title} {Tuning superconductivity in {Ge}:{Ga} using {${\mathrm{Ga}}^{+}$} implantation energy,},}\ }\href {\doibase 10.1103/PhysRevMaterials.5.064802} {\bibfield  {journal} {\bibinfo  {journal} {Physical Review Materials}\ }\textbf {\bibinfo {volume} {5}},\ \bibinfo {pages} {064802} (\bibinfo {year} {2021})},\ \bibinfo {note} {publisher: American Physical
  Society}\BibitemShut {NoStop}%
\bibitem [{\citenamefont {Sacépé}\ \emph {et~al.}(2011)\citenamefont {Sacépé}, \citenamefont {Dubouchet}, \citenamefont {Chapelier}, \citenamefont {Sanquer}, \citenamefont {Ovadia}, \citenamefont {Shahar}, \citenamefont {Feigel’man},\ and\ \citenamefont {Ioffe}}]{sacepe_localization_2011}%
  \BibitemOpen
  \bibfield  {author} {\bibinfo {author} {\bibfnamefont {B.}~\bibnamefont {Sacépé}}, \bibinfo {author} {\bibfnamefont {T.}~\bibnamefont {Dubouchet}}, \bibinfo {author} {\bibfnamefont {C.}~\bibnamefont {Chapelier}}, \bibinfo {author} {\bibfnamefont {M.}~\bibnamefont {Sanquer}}, \bibinfo {author} {\bibfnamefont {M.}~\bibnamefont {Ovadia}}, \bibinfo {author} {\bibfnamefont {D.}~\bibnamefont {Shahar}}, \bibinfo {author} {\bibfnamefont {M.}~\bibnamefont {Feigel’man}}, \ and\ \bibinfo {author} {\bibfnamefont {L.}~\bibnamefont {Ioffe}},\ }\bibfield  {title} {\enquote {\bibinfo {title} {Localization of preformed {Cooper} pairs in disordered superconductors},}\ }\href {\doibase 10.1038/nphys1892} {\bibfield  {journal} {\bibinfo  {journal} {Nature Physics}\ }\textbf {\bibinfo {volume} {7}},\ \bibinfo {pages} {239--244} (\bibinfo {year} {2011})}\BibitemShut {NoStop}%
\end{thebibliography}%

\renewcommand{\thefigure}{\textbf{S\arabic{figure}}}
\setcounter{figure}{0}
\setcounter{equation}{0} 

\clearpage
\onecolumngrid

\begin{center}
    \textbf{SUPPLEMENTARY MATERIAL}
\end{center}

\vspace{0.5cm}

\noindent
\textbf{Femtosecond Ultraviolet Photoelectron Spectroscopy:} One of the techniques we used to evaluate the sample surface was femtosecond ultraviolet photoelectron spectroscopy (fs-UPS). The fs-UPS technique uses femtosecond pulses generated by an amplified Ti:sapphire laser. These pulses are split into pump and probe components. The probe pulses undergo frequency up-conversion to photon energies ranging from 15 to 40 eV through high harmonic generation in Ar gas. Individual harmonics are then directed onto the sample in a UHV analysis chamber, generating standard UPS spectra. The pulse consists of a light at 1.55eV (pump) and 26.35 eV (probe) and is focused on an area on the sample. The probe pulses generate a standard UPS spectrum. Pumping the sample produces a dense electron-hole population that flattens the Nb/Ge bands, which is observed as a rigid shift of the UPS spectrum and corresponds to the band bending. The shift’s direction differentiates between the conditions of depletion and accumulation.

\begin{figure}[ht!]
\includegraphics[width=0.5\textwidth]{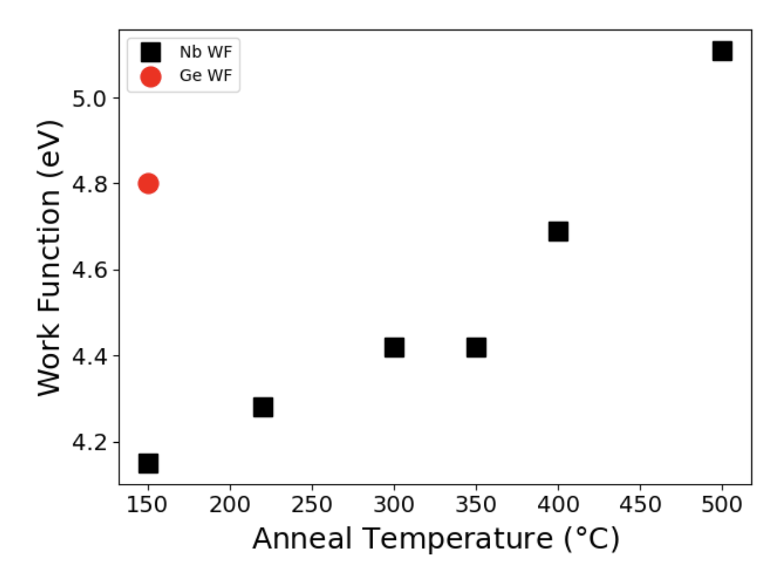}
\caption{\label{fig.S3} The evolution of the work function for an 8 nm thin film of Nb on Ge(001) as a function of annealing temperatures. The increase in the Nb work function with the annealing temperature is consistent with the Ge incorporation into the thin film.
}
\end{figure}

\begin{figure}[ht!]
\includegraphics[width=0.5\textwidth]{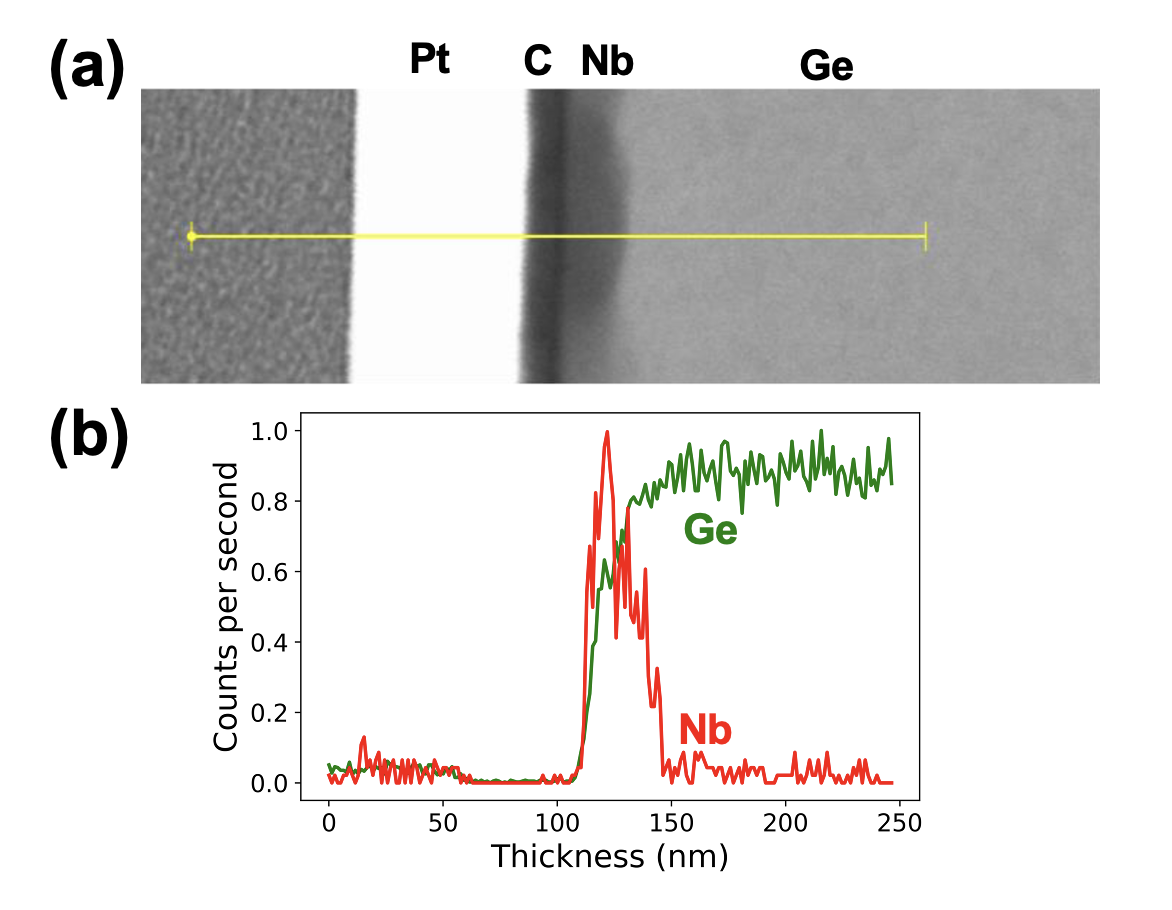}
\caption{\label{fig.S2} (a) Cross-sectional STEM image of an Nb/Ge heterostructure after UHV annealing at 650 $^{\circ}$C. The Nb film is roughened due to the high-temperature annealing. (b) An EDS linescan along the yellow line highlighted in panel (a), shows the normalized ratios for Nb and Ge, including significant incorporation of Ge into the Nb film.
}
\end{figure}

\vspace{0.5cm}
\noindent
\textbf{X-ray photoelectron spectroscopy:} To measure the elemental composition at/near the top surfaces of the Nb/Ge samples we employed X-ray photoelectron spectroscopy (XPS). The XPS was carried out using a PHI VersaProbe III with a monochromatic Al K$\alpha$ X-ray source (hv = 1486.6 eV) with a take-off angle of 45$^{\circ}$. The Al anode was powered at 25 W and 15 kV. The instrument is calibrated to Au and Ag metallic binding energy. Base pressure was above 10$^{-7}$ Torr, and the analysis area size is 1000 x 1000 $\mu$m2 scanned with a beam size of 100 $\mu$m in diameter. The charging compensation was done with dual beam charge neutralization. The binding energies were calibrated to the C 1s peak at 284.8 eV. The survey spectra were collected with a step size of 0.8 eV, dwell time of 50 ms, a pass energy of 224 eV, and three sweeps per spectra. The high-resolution spectra were collected with a pass energy of 69 eV, 0.125 eV step size, and a dwell time of 50 ms, with 3 sweeps. Each sample was sputtered with Ar$^+$ ions at 2 kV acceleration energy over an approximate sputtering area of 3 x 3 mm$^{2}$.

\vspace{0.5cm}
\noindent
\textbf{Atomic Force Microscopy:} We used atomic force microscopy (AFM) to follow the changes to the surface topography of the Nb films throughout the UHV annealing process. The AFM surface topography maps in Fig. S3 shows increased grain sizes with annealing temperature. Additionally, we observe an increase in the root-mean-squared (RMS) roughness at higher annealing temperatures. The as-grown film has an RMS roughness of 4.83 nm. Annealing to 300 $^{\circ}$C, 500 $^{\circ}$C, and 900 $^{\circ}$C increased the RMS roughness to 17.66 nm, 18.56 nm, and 31.22 nm, respectively.

\begin{figure}[ht!]
\includegraphics[width=0.75\textwidth]{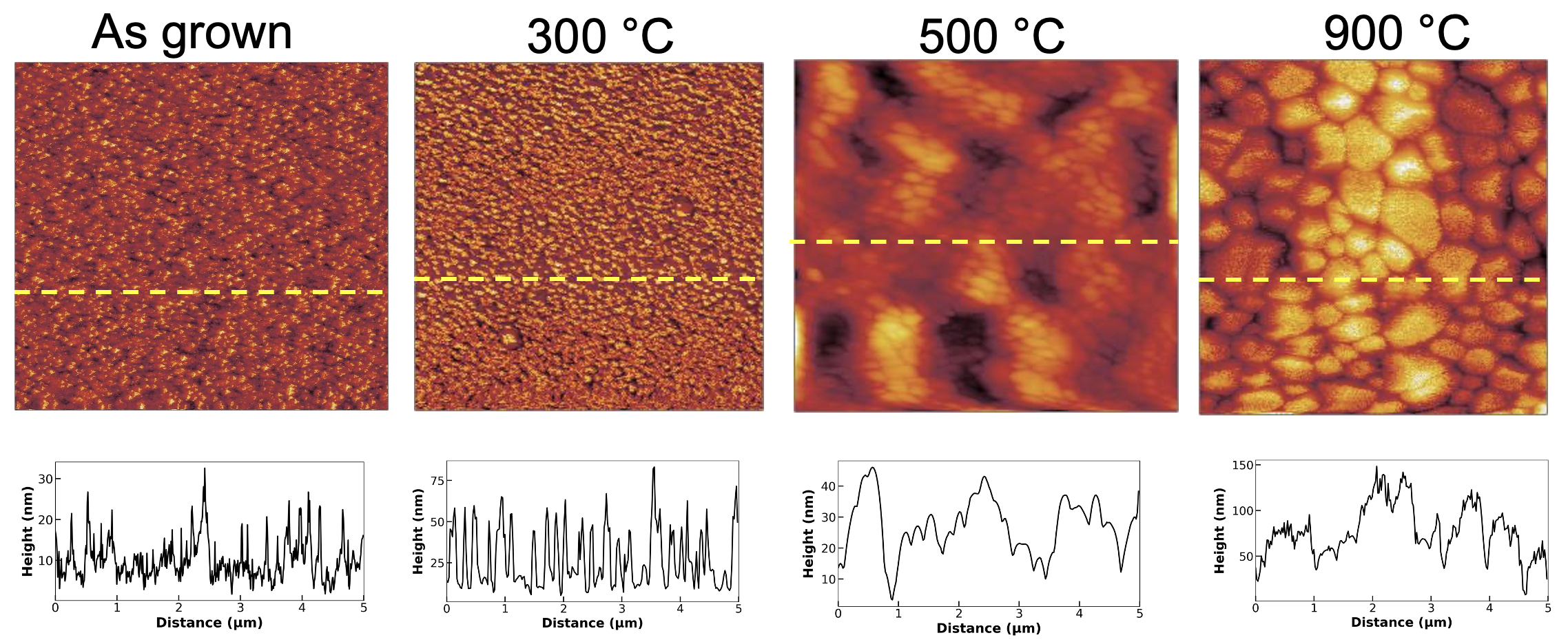}
\caption{\label{fig.S1} Surface topography of Nb/Ge(001) samples as a function of UHV annealing temperature from 300 $^{\circ}$C to 900 $^{\circ}$C. 
}
\end{figure}

\noindent
\textbf{Microfabrication of Nb microwires:} The thin film samples were then fabricated into microwires (widths of 5, 10, and 20 $\mu$m) using electron beam lithography (JEOL 8100 FS E-Beam). The samples were first spin-coated with 495 PMMA A9 at 3000 rpm for 45 sec and baked for 2 minutes at 180 $^{\circ}$C. The e-beam current was set to 40 nA for writing the microwire layout. After e-beam exposure, the resist films were developed in MIBK:IPA 1:3 for 1 minute followed by a 30-s IPA rinse and N$_2$ dry. Samples were then etched using a dry etcher (Oxford Plasmalab System 100 Reactive Ion etching).  A mixture of sulfur hexafluoride (SF6) and argon was used for 1.5 min etch of the Nb layer into the microwire patterns. The residual resist was finally removed by submerging the sample in an acetone bath for 5 minutes, followed by N$_2$ drying.

\begin{figure}[ht!]
\includegraphics[width=0.9\textwidth]{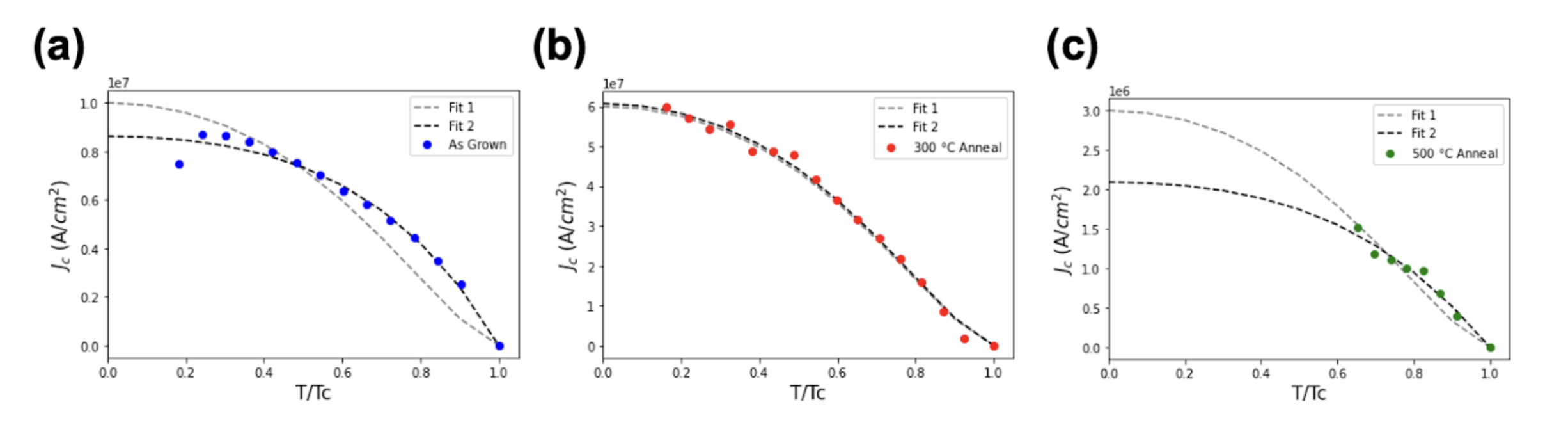}
\caption{\label{fig.S4} Critical current as a function of normalized temperature (t = T/T$_\text{C}$) for three microwires fabricated from 100 nm thick Nb on Ge(001): (a) as-grown Nb/Ge deposited at -170 $^{\circ}$C; (b) UHV annealed at 300 $^{\circ}$C; (c) UHV annealed at 500 $^{\circ}$C. Fit 1 uses the standard equation for the temperature-dependence of critical current in BCS superconductors: J$_\text{C}$ = J$_\text{C}$(0) (1-t$^2$)$^{3/2}$(1+t$^2$)$^{1/2}$. Fit 2 uses a modified version of the Fit 1 formula using the exponent for the (1-t$^2$) term as an additional fitting parameter.
}
\end{figure}

\end{document}